# IS THE MONETARY TRANSMISSION MECHANISM BROKEN? TIME FOR PEOPLE'S QUANTITATIVE EASING

**Sebastian Ilie DRAGOE [1], Camelia OPREAN-STAN[2]**

*Lucian Blaga University of Sibiu, Romania*

***Abstract***

*The monetary transmission channel is disrupted by many factors, especially securitization and liquidity traps. In our study we try to estimate the effect of securitization on the interest elasticity and to identify if a liquidity trap occurred during 1954Q3-2019Q3. The yield curve inversion mechanism shows us that economic cycles are very sensitive to decreasing profitability of banks. However there is no evidence that restoring their profits will ensure a strong recovery. In this regard, we research the low effect of Quantitative Easing (QE) upon economic growth and analyze whether securitization and liquidity traps posed challenges to QE or is it the mainstream theory flawed. In this regard we will examine the main weaknesses of QE, respectively the speculative behavior induced by artificial low rates and its unequal distribution. We propose a new form of QE that will relief households and not reward banks for their risky behavior before recession.*

## 1. Introduction

The monetary transmission channel is disrupted by many factors like securitization, liquidity traps and the threat of crypto-assets. While stable coins and proposals like Libra do not suffer from the high volatility characteristic of crypto assets, due to high default risks, lack of insurance and low transaction volumes, they do not represent a significant threat for the monetary channel. However, securitization and liquidity traps create important difficulties in formulating and implementing monetary policies.

[1] *Ph.D. Student, Faculty of Economic Sciences, Lucian Blaga University of Sibiu, Calea Dumbrăvii nr.17, 550324, Sibiu, Romania, e-mail: dragoesebastian@yahoo.com*

[2] *Professor Ph.D., Faculty of Economic Sciences, Department of Finance-Accounting, Lucian Blaga University of Sibiu, Calea Dumbrăvii nr.17, 550324, Sibiu, Romania, e-mail: camelia.oprean@ulbsibiu.ro*

In the last four decades, the financial system of United States experienced an era of financial innovation and nonbank super-financialization. It would look like the nonbanking sector has grown at much a higher pace the private deposit institutions. However, if we take into account the securitization process the picture could be quite different because banks and the other private depository institutions have used securitization in order to elude monetary policy tightening. Banks used Issuers of Asset Backed Securities to convert loans into securities. By selling loan portfolios to IABS, private depository institutions will acquire liquidity, meaning that it doesn't have to wait for the loans to be repaid or to obtain funds from central bank in order to have liquid assets.

The other important factor that hinders the effectiveness of monetary policy is the liquidity trap via Hicks and Krugman style. The definition provided by Krugman (1998) is different than Keynes' (1936). The original liquidity trap states that at low interest rates the economic agents could prefer cash instead of bonds and monetary policy will lose control of interest rate. Krugman's liquidity trap states that at lower or 0% interest rate, conventional monetary policy tools can no longer stimulate economy as the natural interest rate is negative. This concept is closely related to zero lower bound. The last type of liquidity trap is believed to have taken place during the Great Recession.

The purpose of this paper is to estimate the effect of monetary policy on GDP growth and to research the main disruptive factors of monetary policy, namely securitization and liquidity traps. The research methods used in this paper are especially quantitative analysis (OLS regressions) and deductive methods. The main contributions of the article consist in estimating the effects of the monetary policies upon economic activity and ways to enhance the impact of monetary policy on changes in real GDP.

The reminder of the paper is structured as follows: in Section 2, namely the literature review we present the research regarding the impact of securitization on interest elasticity and the article is describing the liquidity trap. In the methodology section, we highlight the econometric models used in the paper. Next, in the research findings, we are presenting the results and implications of the econometric models, and in the final section we conclude.

## 2. Literature review

Estrella (2002) has researched the impact on output gap of interest rates, when allowed to separately vary with securitization of single and multi-family home mortgages, by extending the IS regression of Rudebusch and Svensson (1999). His conclusion was that monetary policy has lost its direct effect on output gap. However, the decline of output gap' sensitivity to changes in interest rates depends on liquidity and on the supply of credit as changes in the federal funds rate are transmitted more directly to the mortgage rates. Long *et al.* (2009) reached to the same conclusion and with the help of a VECM model estimated that the increase of securitization diminished the total effect of interest rate shocks on output gap by over 90%.

The effect of liquidity trap on the efficacy of monetary policy in the updated form has been pioneered by Hicks (1937) and Krugman (1998). Unlike the original liquidity trap which stated that at very low nominal interest rates central banks lose the control over interest rates, the updated liquidity trap claims that monetary policy loses effectiveness. Bernanke (2000) described Japan's liquidity trap as a self-induced paralysis by the central bank arguing that central banks have many tools even at zero lower bound. Krugman and Eggertsson (2011) formalized a model of a liquidity trap produced by a deleveraging shock (debt-deflation). Basically all authors claim that currency devaluation, higher inflation target, and a very large expansion of monetary base (quantitative easing) can end the liquidity trap. However, their results are debatable in the sense that despite the 0% interest rate and the large-scale asset purchases by Federal Reserve, money supply (M2), inflation, and economic growth have not paced up during quantitative easing. Benanke (2009) uses the term "credit easing" because Federal Reserve also focused on the way the composition of its balance sheet affects credit conditions. As European Central Bank also used mixed programs (including purchases of corporate bonds starting in 2016) without significant effects on economic growth, the reasons why United States performed better than Eurozone and Japan are independent of the distinction between the adopted monetary policies. As a result credit easing should have better but small and negligible effects on credit conditions when compared with quantitative easing. This is one of the reasons why we will use the term "quantitative easing" alongside the fact that we are researching the expansion of Federal Reserve's balance sheet, not its composition.

The identified knowledge gap in the existing literature is the lack of studies regarding the distribution effects of quantitative easing and how it affects the efficiency of this policy measures. As the banks are borrowing cheaper and are bailed-out, QE might have just distributional effects rather than a strong impact on economic growth.

## 3. Data and methodology

In order to estimate the extent of securitization and the share of loans originating from private depository institutions we follow a similar approach to Unger (2016).

We use data for private depository institutions, for shadow banks (Finance Companies –we use FC as abbreviation, Security Brokers and Dealers – SBD and Issuers of Asset Backed Securities - IABS) and Government Sponsored Enterprises (GSE). We calculate the origination of credit for period sample 1968Q1-2019Q1. Starting date is 1968 as this is the first year of mortgage-backed securities issuance, McConnell and Buser (2011), although IABS, on which we focus, have started their activity in 1984 Q4. The methodologies are distinct in the sense that we use aggregate loans for each type of institution (GSELA, IABSLA, SBDOLAA, FCLA and FL704023005.Q) and have a different allocation into classes. While (Unger, 2016) has split the origination and the holding of loans into traditional banking (private depository institutions + Government Sponsored Enterprises) and shadow banking, we divide them into private depository and non-private depository institutions (FC, SBD, IABS and GSE). We compute the loans that originate from private depository by multiplying its share of holdings in total loan holdings to the holdings of IABS. The sum of the resulted amount and the private depository's holdings is divided by total loan holdings in order to obtain the share of loans originating from private depository institutions.

We compute the securitization ratio for private depository institutions as a fraction with the IABS loans originating from private depository institutions as numerator and the sum of balance sheet loans and securitized loans provided by the depository institutions as the denominator.

We estimate the impact of securitization on the monetary transmission mechanism with the help of 2 regression models for period 1959Q1-2019Q1. Real economic growth (change after 4 quarters) is the dependent variable in both models. Real GDP growth is regressed in the first equation on economic growth with one lag (expressed in logarithm), on economic growth with two lags and on moving federal funds rate (average of the current and last 3 quarters), also with one lag (see equation 1). In the second regression we allow the interest elasticity of output to vary with the securitization ratio (see equation 2). The models used resemble the models developed by Estrella (2002). Estrella (2002) models research the impact of single family and multi-family securitized mortgage on the real interest elasticity of Output gap while our models analyze the effect of the securitization of credit originating from private depository institutions on the nominal interest of real GDP growth.

$$\Delta Y = \alpha + \beta_1 \Delta Y_{t-1} + \beta_2 \Delta Y_{t-2} + \beta_3 \Delta \bar{I}_{t-1} \quad (1),$$

where Y is real GDP and Ī represents average federal funds rate.

$$\Delta Y = \alpha + \beta_1 \Delta Y_{t-1} + \beta_2 \Delta Y_{t-2} + (\beta_3 + \beta_4 Sr)(\Delta \bar{I}_{t-1}) \quad (2),$$

where Sr is securitization ratio.

For identifying the new form of liquidity trap described by Krugman we will employ Taylor rule for 1954Q3-2019Q3.

Taylor rule is an instrument used by the policymakers in order to harness inflation and to close the output gap. The formula suggested by Christian Zimmermann (2014) and used by St Fed Louis is:

$$GDP\ Deflator + 0.02 + 0.5 * (GDP\ Deflator - 0.02) + \frac{0.5*(Real\ GDP - Real\ Potential\ GDP)}{Real\ Potential\ GDP} \quad (3)$$

If the suggested nominal interest rate is negative it means that the economy is in a liquidity trap.

As we want to check if the monetary stimulus, especially the monetary base had any effect on economic activity we will construct chain index for monetary base, money stock (M2) and nominal GDP for 2007Q3 (the first quarter before the Great Recession) -2019Q1. As the time series for M2 and GDP are seasonally adjusted, we have used Tramo/Seats procedure. Both programs, Tramo and Seats are used together. TRAMO was developed by Gómez and Maravall (1992) is utilized for estimating, forecasting and for the interpolation of regression models which are lacking observations and with ARIMA errors. SEATS decomposes the chosen time series into its trend, cycle, seasonally adjusted and irregular components by using a model based on the ARIMA method.

## 4. Research Findings

In terms of loan origination, the share of private depository loans has not decreased so dramatically before the Great Recession (see figure 1 and 2). After the Great Recession, the process reversed and both in terms of origination and holdings, the private depository institutions dominate.

In 2010Q1 a sharp peak is observable in loan origination of private depository entities and in the loan holdings of non-banks. This is caused by the Freddie Mac and Fannie Mae mortgage pools which were moved back into their balance sheets due to new accounting rules.

**Figure 1. The composition of loan holdings**

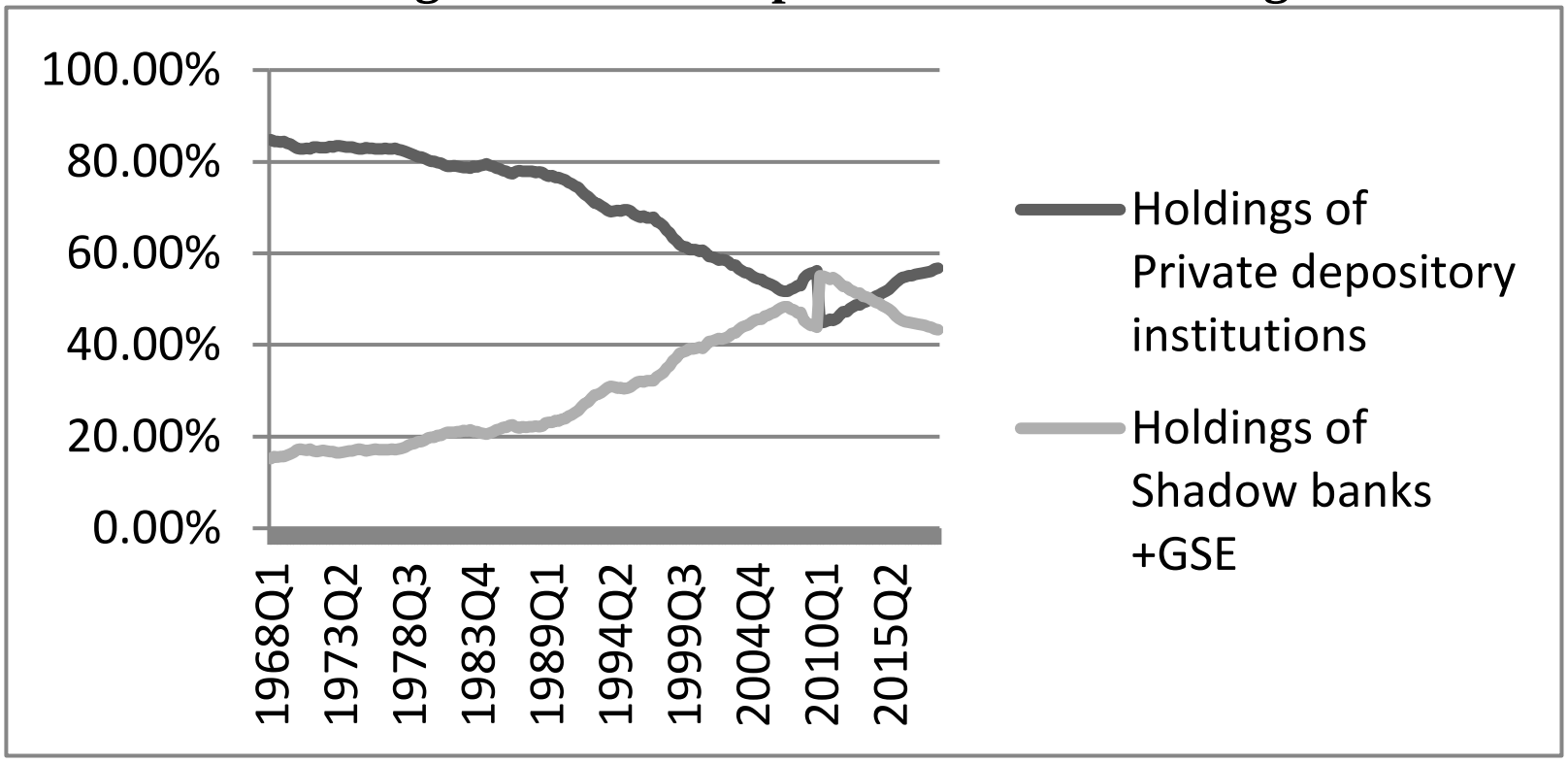

Source: Authors' calculations

**Figure 2. The composition of loan origination**

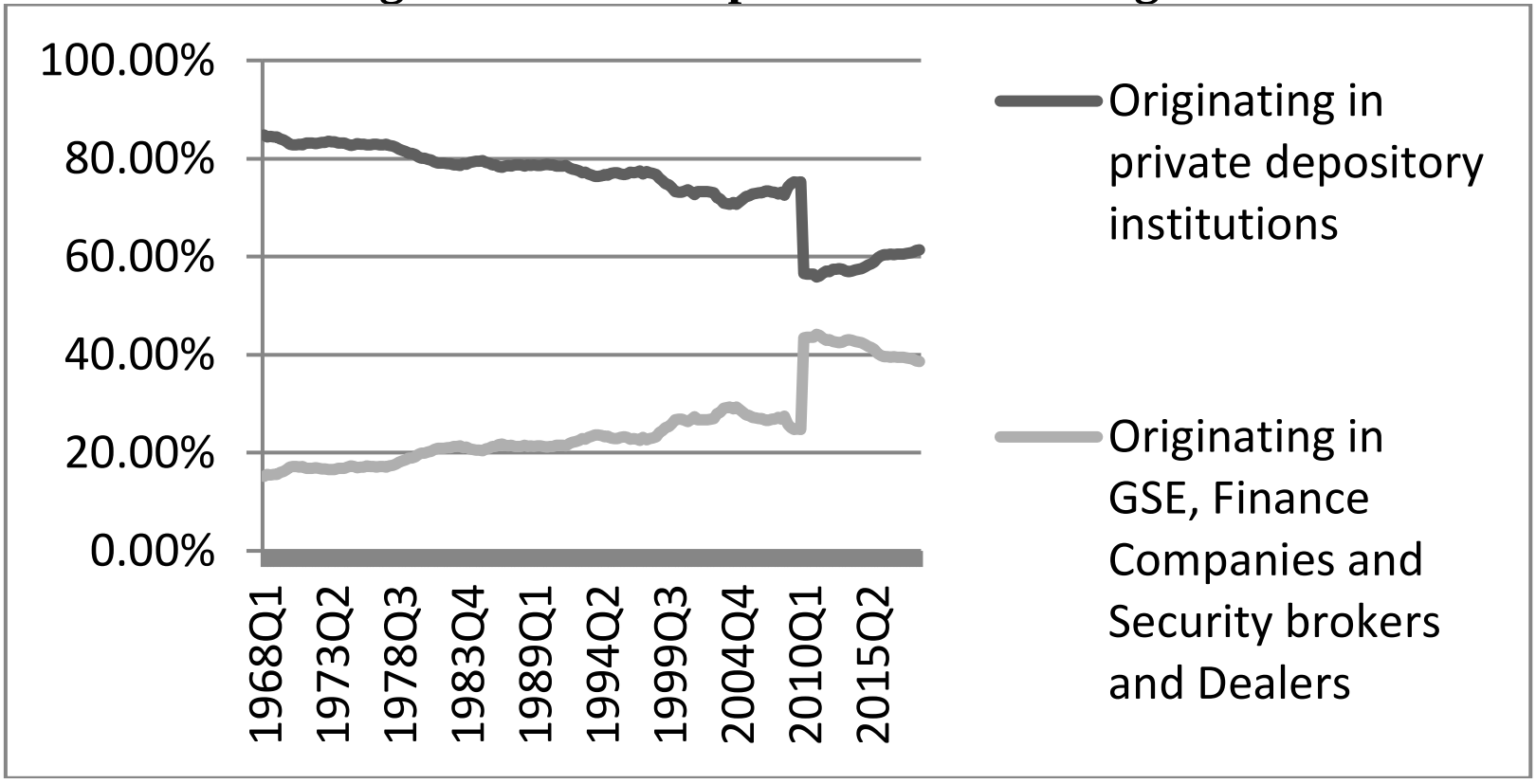

Source: Authors' calculations

The highest value of the securitization ratio is almost 29.34% and it was registered in 2007Q3 (see figure 2), meaning that the securitized loans represented 41.51% of balance sheet loans.

**Figure 3. Securitization ratio**

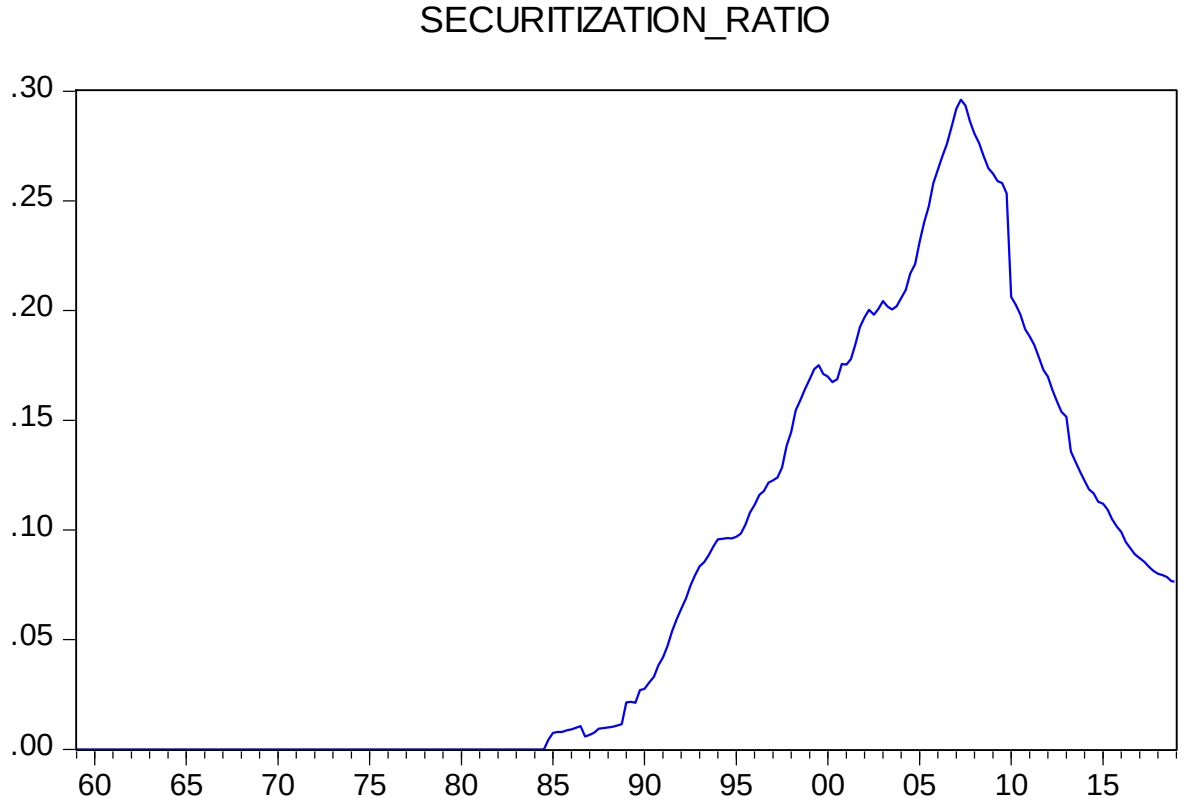

Source: Authors' calculations

The next step is to check how potent monetary policy is nowadays, considering the fact that many banks resort to securitization in order to avoid monetary policy tightening and capital requirements and to acquire fast funding.

**Figure 4.Monetary policy impact on real GDP growth**

Dependent Variable: L_RGDP-L_RGDP(-4)
Method: Least Squares
Date: 12/26/19 Time: 00:10
Sample (adjusted): 1960Q3 2019Q1
Included observations: 235 after adjustments

| Variable | Coefficient | Std. Error | t-Statistic | Prob. |
|---|---|---|---|---|
| C | 0.004084 | 0.001196 | 3.413685 | 0.0008 |
| L_RGDP(-1)-L_RGDP(-5) | 1.219121 | 0.058910 | 20.69480 | 0.0000 |
| L_RGDP(-2)-L_RGDP(-6) | -0.354594 | 0.061653 | -5.751443 | 0.0000 |
| D(L_FED_FUNDS_AVERAGE(-1)) | -0.011639 | 0.005178 | -2.247839 | 0.0255 |

| | | | |
|---|---|---|---|
| R-squared | 0.810669 | Mean dependent var | 0.029887 |
| Adjusted R-squared | 0.808210 | S.D. dependent var | 0.021887 |
| S.E. of regression | 0.009585 | Akaike info criterion | -6.440343 |
| Sum squared resid | 0.021223 | Schwarz criterion | -6.381456 |
| Log likelihood | 760.7403 | Hannan-Quinn criter. | -6.416602 |
| F-statistic | 329.6941 | Durbin-Watson stat | 2.110635 |
| Prob(F-statistic) | 0.000000 | | |

Source: Authors' calculations

The p-values associated with the model and with the coefficients are under the selected level of significance of 5%, the model is valid. The high coefficient of determination is more than 0.81 imply that the selected variables have explanatory power. Although, the p-value associated to the coefficient of the changes in interest rates is less than 5%, the value of the coefficient is very close to 0. We can deduce that changes in interest rates have small inverse effects on economic growth.

**Figure 5. Recursive coefficient of interest elasticity**

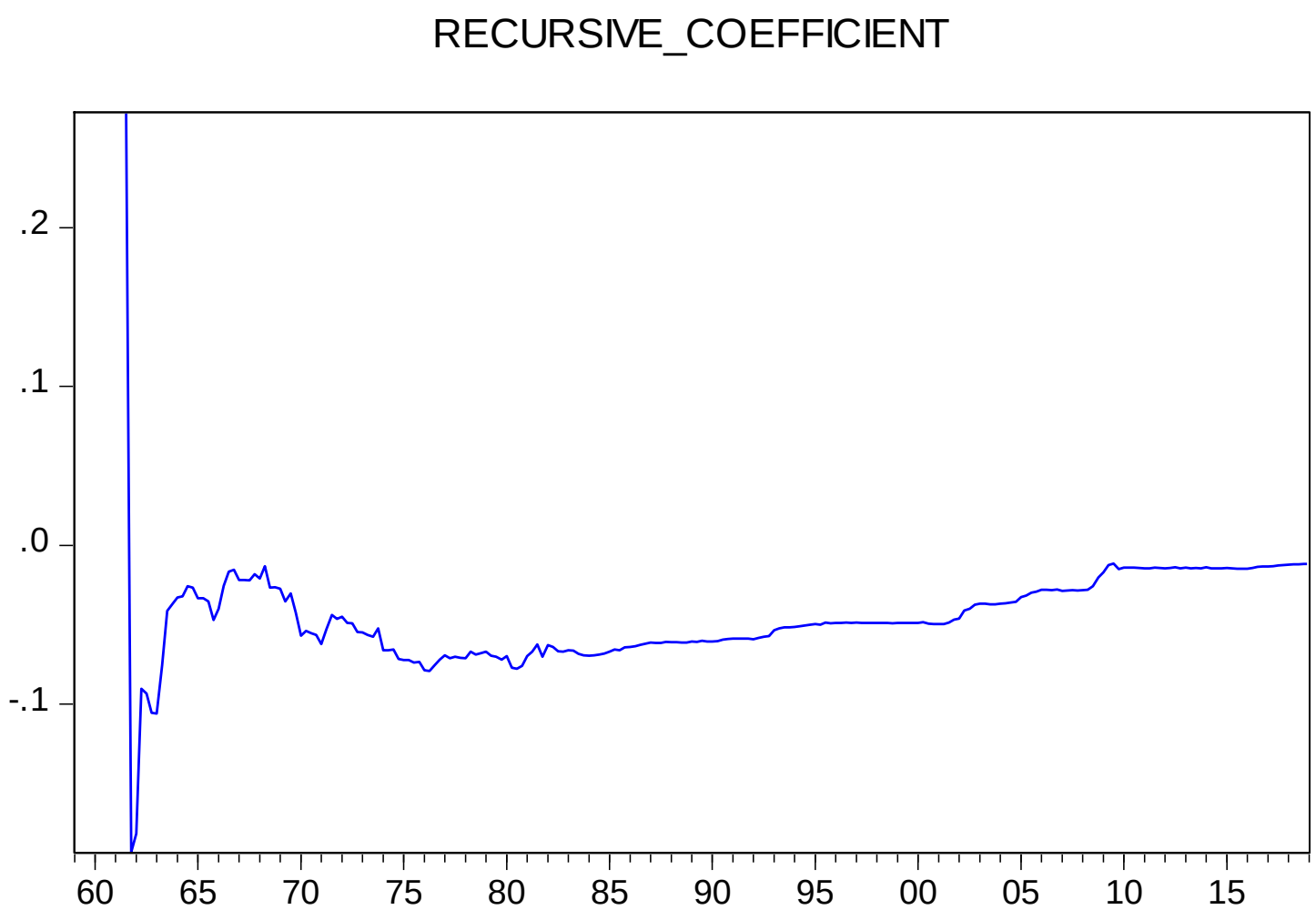

Source: Authors' calculations

The recursive coefficient shows that the impact of interest rates on real GDP growth since the start of securitization through IABS. Although the securitization level peaked in 2007Q3 and afterwards followed a downtrend, the effect of monetary policy on economic activity continued to decline until 2009Q3 and after it stabilized suggesting there is at least one more factor. We will come back later after estimating the interest elasticity when allowed to vary with the securitization ratio (see figure 6).

**Figure 6. Monetary policy with the degree of securitization's impact on real GDP growth**

Dependent Variable: L_RGDP-L_RGDP(-4)
Method: Least Squares
Date: 12/26/19 Time: 00:10
Sample (adjusted): 1960Q3 2019Q1
Included observations: 235 after adjustments
L_RGDP-L_RGDP(-4)=C(1)+C(2)*(L_RGDP(-1)-L_RGDP(-5))+C(3)
*(L_RGDP(-2)-L_RGDP(-6))+(C(4)+C(5)*D(SECURITIZATION_RATIO))
* D(L_FED_FUNDS_AVERAGE(-1))

| | Coefficient | Std. Error | t-Statistic | Prob. |
|---|---|---|---|---|
| C(1) | 0.004147 | 0.001201 | 3.453304 | 0.0007 |
| C(2) | 1.224265 | 0.059415 | 20.60535 | 0.0000 |
| C(3) | -0.360483 | 0.062272 | -5.788803 | 0.0000 |
| C(4) | -0.012606 | 0.005359 | -2.352354 | 0.0195 |
| C(5) | -0.305969 | 0.430351 | -0.710977 | 0.4778 |

| | | | |
|---|---|---|---|
| R-squared | 0.811084 | Mean dependent var | 0.029887 |
| Adjusted R-squared | 0.807798 | S.D. dependent var | 0.021887 |
| S.E. of regression | 0.009595 | Akaike info criterion | -6.434027 |
| Sum squared resid | 0.021176 | Schwarz criterion | -6.360419 |
| Log likelihood | 760.9982 | Hannan-Quinn criter. | -6.404352 |
| F-statistic | 246.8676 | Durbin-Watson stat | 2.117533 |
| Prob(F-statistic) | 0.000000 | | |

**Source: Authors' calculations**

The outcome is not surprising. Securitization has a considerable effect on the potency of monetary policy upon economic growth. When we allow the interest elasticity to vary with the weight of securitized loans we obtain a higher absolute coefficient, equal to -0.306 (see figure 23) however since the p-value is 0.4778, we can deduce that monetary policy has lost its direct effect on economic activity. Due to securitization, the depository institutions with constrained liquidity no longer required cheaper funding and/or greater volumes of monetary base from central banks for extending loans.

Even though securitization has adversely affected the transmission of monetary policy especially in the case of monetary tightening, Federal Reserve used this channel in order to stabilize the economy during the Great Recession through programs like TALF, Q1, Q2 and Q3.

However, despite the downward trend and the usage of this channel for stimulus, the interest rates continued to lose direct effect on economic growth until 2009Q3 inclusively, indicating that the monetary mechanism was harmed during the Great Recession and the cause might be a liquidity trap.

**Figure 7. Federal Funds Rate and Taylor Rule (1954Q3-2019Q3)**

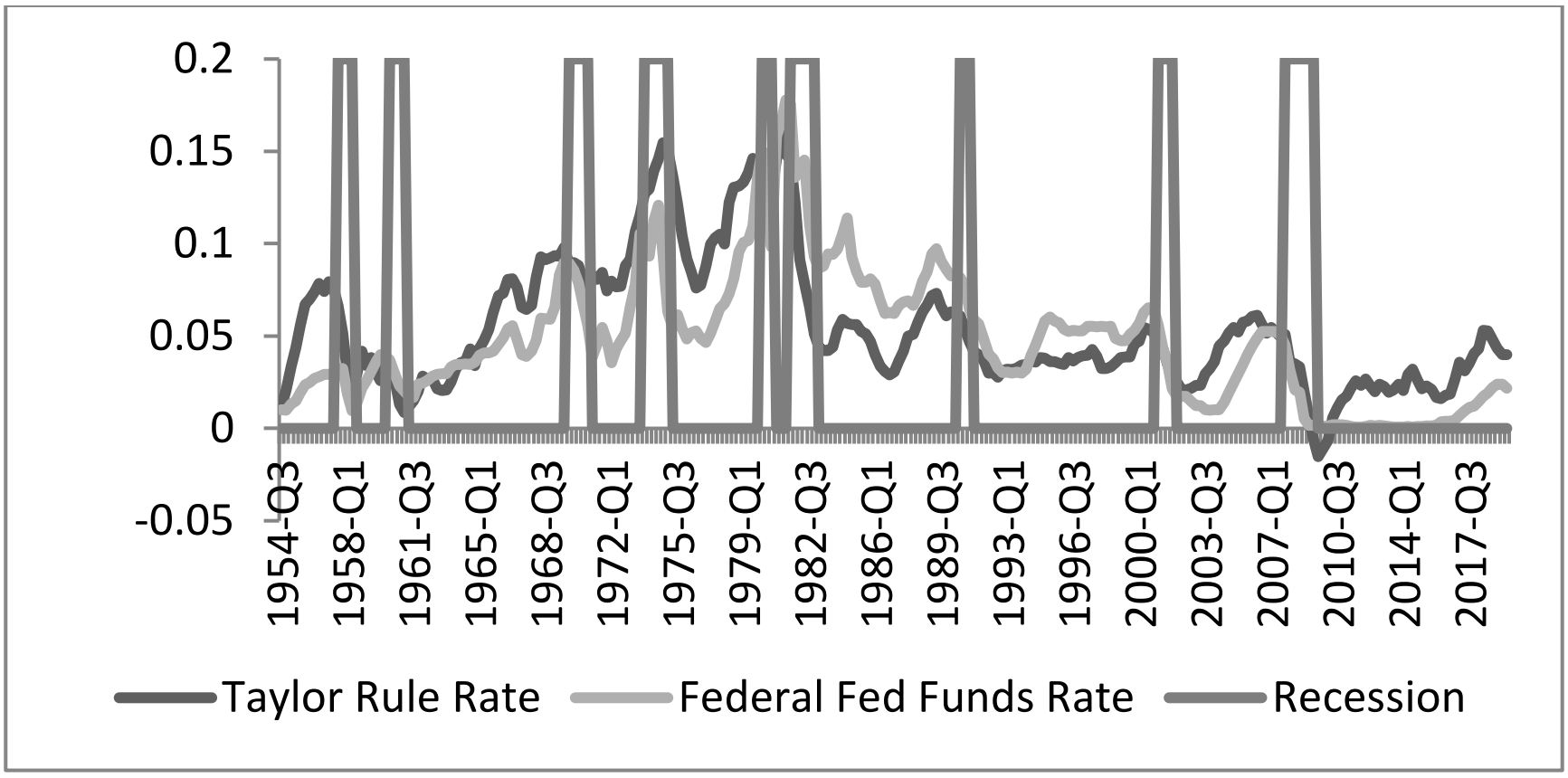

Source: data from St Fed Louis

As shown in figure 7, Taylor rule rate became negative during 2009Q2-2010Q1, indicating that a Krugman style liquidity trap occurred. For compensating the zero lower bound, Federal Reserve has maintained an almost 0% interest rate until 2015Q4 and used quantitative easing. In doing so Federal Reserve helped banks in restoring the banks' profitability. We know that financial institutions profitability can cause recessions because the yield curve inversions have preceded the last 7 recessions in United States (Haltom *et al*, 2018). The yield curve inversion is such a good predictor because it indicates a tight monetary policy (Koenig and Phillips, 2019),an uncertain economic outlook, decreasing profitability relative to costs for banks and upcoming stricter credit standards and a possible credit crunch (Wheelock, 2018). Indeed, decreasing financial institutions' profitability can produce recessions but is the vice versa also true? Can restoring financial institutions' profitability induce a strong recovery? The evidence shows that we can reject this hypothesis. Not only interest elasticity coefficient is very low, but QE also seems to have a low effect on economic activity.

QE started in November 2008 and ended on October 2014. As we can see in figure 8, the impact of QE on money supply and nominal GDP was very weak (see figure 8). Martin Feldstein (2015) argued that the Federal Reserve's decision to pay interest for excess reserves. However, Bank of Japan introduced Complementary Deposit Facility only in 2008. During 2001 –2005, Bank of Japan almost doubled the monetary base with little effects on inflation and economic growth, without paying interest on excess reserves. The other mainstream theorists who are in favor of using monetary policy are hiding under the umbrella of the new form of liquidity trap, although liquidity trap did not last so much as QE. In fact monetary base expansion can't increase money supply if banks don't have appetite for risk. Also in practice there are more interest rates. Central banks can only control the front end of the yield curve with the short-term exception of old-fashioned liquidity trap which imply cash hoardings and at interbank level a spike in the differential between federal funds rate and treasury rate, with both processes setting a temporary floor rate for their corresponding level. Central banks can influence the long-term interest rates but can't control them as the long-term interest rates also reflect the investor expectations.

**Figure 8. Monetary base, money supply and Nominal GDP evolution**

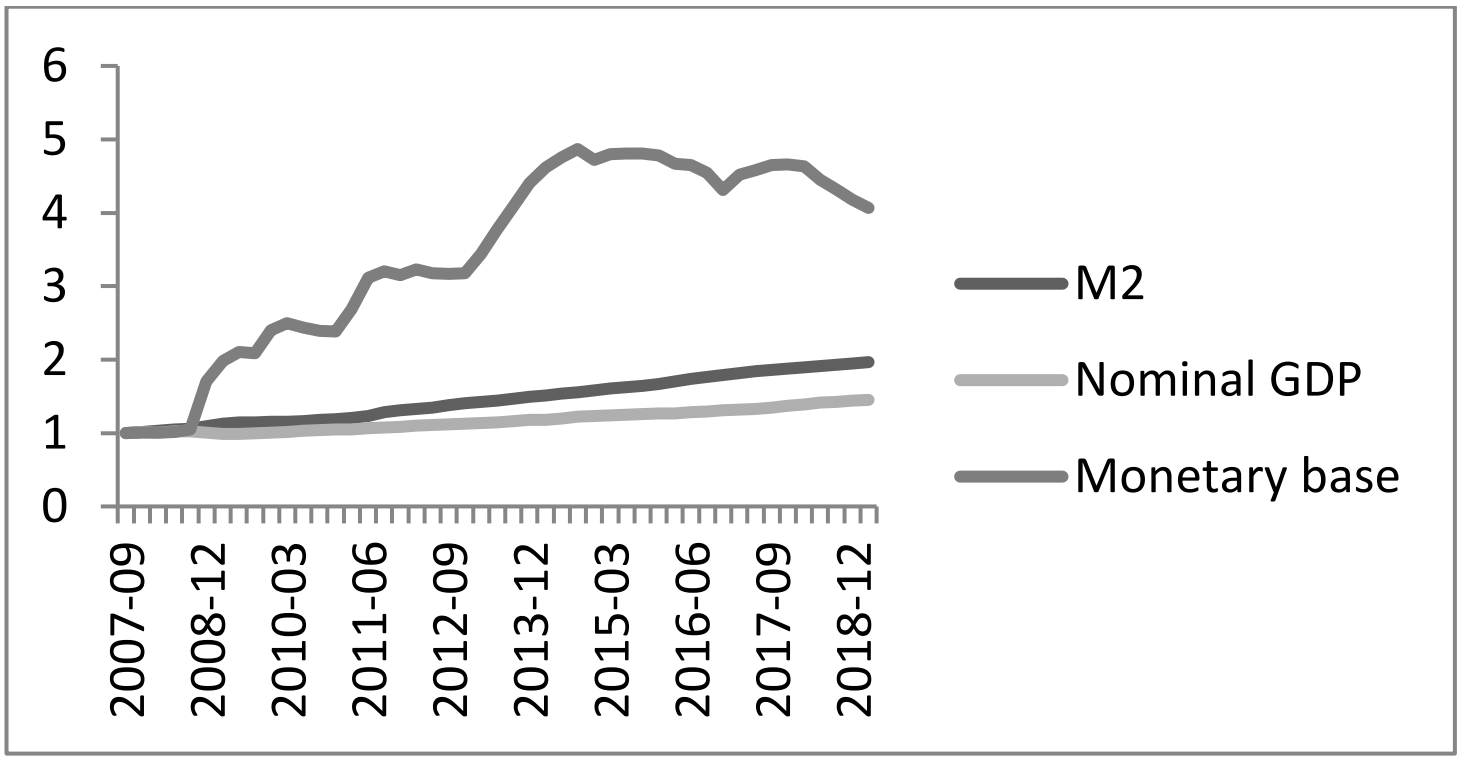

Source: Authors' calculations, data from BEA and Federal Reserve, tables H3 and H6

Thus banks' low risk strategies are to blame for the ineffective monetary policy during quantitative easing period. But in figure 8 we can observe another event: Nominal GDP increased at a slower pace than money supply (M2). This implies a decrease in money velocity, which means that more and more money are being used for speculations. The speculation motive is the most sensitive to artificial low interest rates. When interest rates are close to 0 and the economy is no longer in the recession phase, investors can seek riskier assets with higher yields and use leverage. As we can see in figure 9 money velocity and the velocity of domestic nonfinancial debt securities decreased. However, the Velocity of total domestic debt securities increased meaning that re-intermediation shrunk. The highest level of debt securities and loans for domestic financial sector was recorded in 2008Q3. Besides the speculative behavior, the over reliance on monetary policy can also lead to unequal distribution. Hayek, the Fed's expansion of monetary base is not distributed uniformly ”like water into a tank, but rather oozes like honey into a saucer, dolloping one area first and only then very slowly dribbling to the rest” (Spitznagel, 2012). Fed’s helicopter money means cheaper funding for the largest banks which will be converted into capital transfers.

**Figure 9. Money and credit velocities**

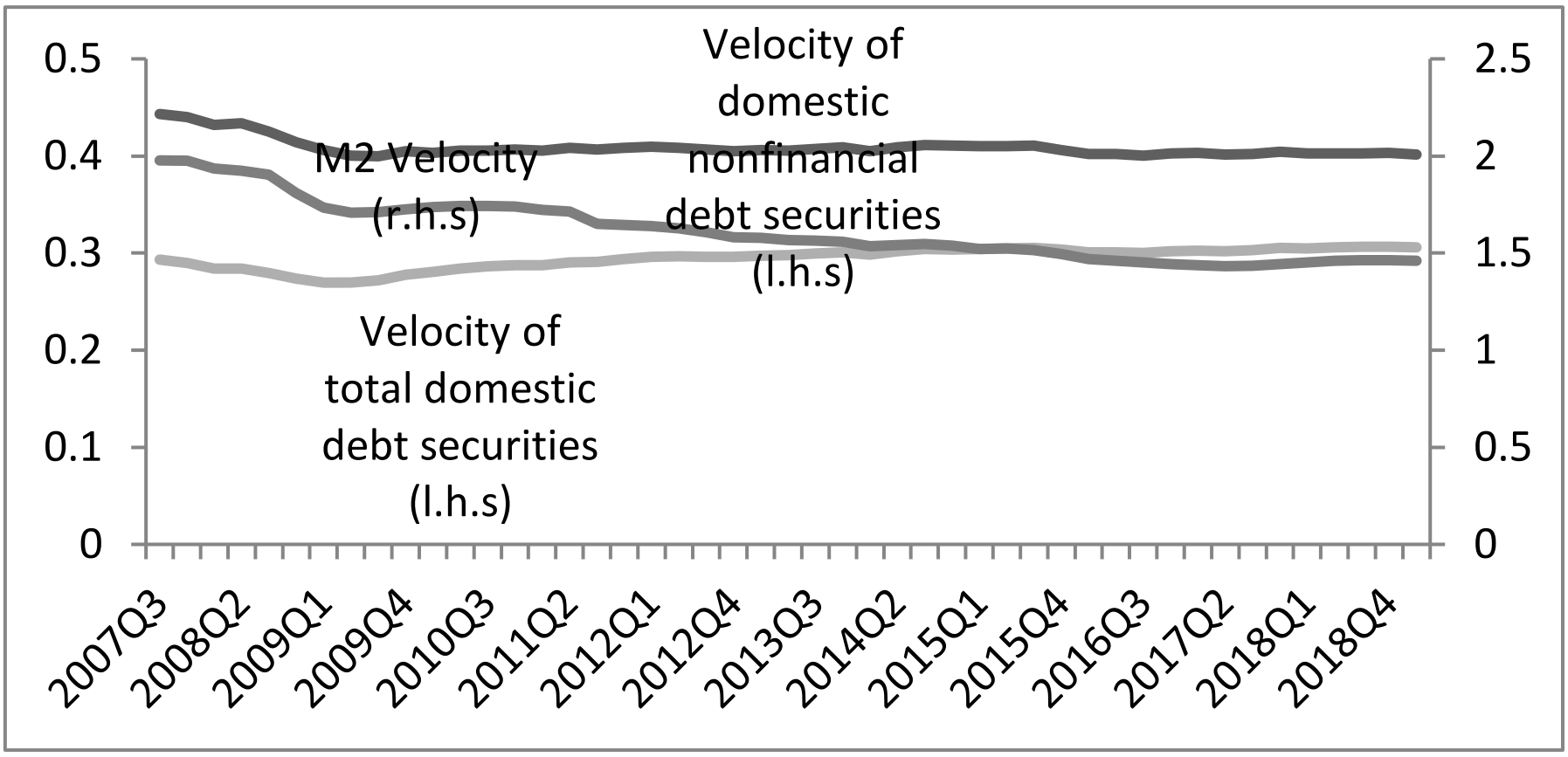

**Source: Authors’ calculations**

It seems that Federal Reserve has no ammunition left to fight the current recession produced by Coronavirus. Interest rates are already very low: the effective federal funds rate was in average 0.65% in March 2020 and United States will most likely fall into another liquidity trap. Also, the securitization rate is much lower than the registered level before the Great Recession. In this context TALF 2.0 and the other asset backed securities oriented programs will have a lesser impact on economy.

Instead of rewarding banks and corporations for incompetence and for lack of precautionary measures like high liquidity and high private inventories for factories, we propose peoples' QE without reimbursement. This is also part of the 2 trillion $ stimulus named Cares Act. As this assistance will be provided to low income families, the effect will be higher, as permanent income theory assumes that transitory incomes will have a reduced effect on consumption. Also is more probable that the trade deficit would not explode since during crisis the openness is decreasing. However it is financed with public debt.

A better alternative to expanding the public debt would be for Fed to relief households especially during difficult times like the current lockdown. This solution would solve the distribution problem evocated by Hayek and will be more efficient due to the lack of intermediaries. The fact that this measure is inflationary is not a flaw. Actually, reflation and stagflation are the only ways an economy can avoid a depression. Due to moderate or high inflation, the cash flows are rising and enable the debt repayment of the debt contracted during boom (Keen, 2011).

## 5. Conclusions

The results of our analysis proved that interest elasticity coefficients are very small and when we take into account the securitization, interest rate has lost its direct effect on output. Thus, the changes in monetary policy have a low impact on economic growth. This is in line with the other studies from the existing literature, namely Estrella (2002) and Long *et al.* (2009), although in comparison with the mentioned studies, we have estimated the effects of nominal interest rates, not of the real interest rates.

The other very important factor, the Krugman style liquidity trap was identified during 2009Q2-2010Q1, as the Taylor rule rate was negative. Policymakers attempted to compensate the implied negative neutral interest rate with QE. The effect on money supply, inflation and economic growth was reduced. Not only the transmission mechanism was damaged, but in fact monetary base expansion does not have the intended effects if the banks don't resume lending and seek safer investments.

There are also other weaknesses for monetary policy especially at 0% interest rate, respectively the speculative behavior and the unequal distribution.

Our recommendation for Federal Reserve is to eliminate these flaws by funding directly the households without reimbursement. The package would be more effective if it will be financed by Federal Reserve instead of using governmental debt.

As possible directions for further research, we aim to investigate the impact of the liquidity trap on the long term economic growth and to estimate the money multiplier effect of the quantitative easing with and without the extent of excess reserves.